\documentstyle[12pt]{article}
\newcommand{\CQG}{{\em Class. Quantum Grav.} }
\newcommand{\AP}{{\em Ann. Phys., Lpz.} }
\newcommand{\JPSJ}{{\em J. Phys. Soc. Japan\/} }
\newcounter{llista}
\newcounter{llista1}
\newtheorem{theorem}{Theorem}
\newtheorem{lemma}{Lemma}
\newtheorem{proposition}{Proposition}

\newtheorem{definition}{Definition}

\begin{document}
\author{J Llosa and D Soler}
\title{Fermat-holonomic congruences}
\date{9th March 2000}
\maketitle

\begin{abstract}
Fermat-holonomic congruences are proposed as a weaker substitute for the too restrictive class of Born-rigid motions. The definition is expressed as a set of differential equations. Integrability conditions and Cauchy data are studied.
\end{abstract}

\section{Introduction}

The relevance of rigid motions in Newtonian mechanics basically stems from
the following facts: (i) they model the motions of {\it ideal} rigid bodies,
and also the behaviour of {\it real} rigid bodies at the first approximation
level, (ii) they provide a definition of strains which, in elasticity
theory, determine stresses, and (iii) they describe the relative motion of
two Newtonian reference frames, i.e., those frames where Newton's laws of
mechanics hold, provided that inertial forces are taken into account.

A common feature of Newtonian rigid motions is that each one is
unambiguously determined by the giving of the trajectory of one point
together with the motion's vorticity along that line (that is, the angular
velocity).

The relativistic generalization of rigid motions needs to be formulated in
terms of a spacetime manifold (${\cal V}_4$, g). A {\it motion }is then
defined by a 3-parameter congruence of timelike worldlines, ${\cal C}$, 
\begin{equation}  \label{E.2}
x^\alpha (t)=\varphi ^\alpha (t,\,y^1,\,y^2,\,y^3)
\end{equation}
where $\,y^1,\,y^2,\,y^3$ are the parameters. In its turn, the congruence $%
{\cal C}$ is determined by its unit timelike velocity field 
\begin{equation}  \label{E.2a}
u^\alpha (x)\,,\quad \mbox{ with} \qquad g_{\mu \nu }u^\mu u^\nu =-1
\end{equation}

The stationary space ${\cal E}_3$ for the congruence is the quotient space,
where cosets are worldlines in ${\cal C}$. The Fermat tensor \cite{BEL90} 
\begin{equation}  \label{E.3}
{\hat g_{\alpha \beta }:=g_{\alpha \beta }+u_{\alpha}u_{\beta}}
\end{equation}
yields the infinitesimal radar distance 
\begin{equation}  \label{E.3a}
d\widehat{l}_R^2=\widehat{g}_{\mu \nu }(x)dx^\mu dx^\nu
\end{equation}
between two neighbour worldlines. This quantity is not usually constant
along a worldline in ${\cal C}$. Hence, it does not define an infinitesimal
distance on ${\cal E}_3$.

Only in case that the Born-rigidity condition \cite{BORN09} 
\begin{equation}  \label{E.4}
\Sigma_{\alpha\beta} = {\cal L}(u)\,\hat g_{\alpha\beta} = 0
\end{equation}
holds, $g_{\alpha \beta }$ defines a Riemannian metric on ${\cal E}_3$.

The above condition (\ref{E.4}) consists of six independent first order
partial differential equations with three independent unknowns, namely, $u^i$%
, $i=1,2,3$ (since $u^4$ can be obtained from (\ref{E.2a})), just like in
the Newtonian case.

The class of Born-rigid motions would generalize Newtonian rigid motions
also because some spatial distance between points in space is conserved.
Unfortunately, the Herglotz-Noether theorem \cite{HER-NOE10} states that,
even in Minkowski spacetime, the class of Born-rigid motions is narrower
than sought. Indeed, motions combining arbitrary acceleration and rotation
are excluded from this class.

Nevertheless, this shortness should not be surprising. Indeed, six first
order partial differential equations for only three unknown functions
unavoidably entail integrability conditions, which yield additional
equations. The latter will lead to new integrability conditions, and so on.
The process of completing the partial differential system (\ref{E.4}) ends
up with a set of equations which is too restrictive for our expectations
(namely, six degrees of freedom or six arbitrary functions of time).

In a recent work \cite{LLOSA97} by one of us, 2-parameter congruences in a
(2+1)-dimensional spacetime, ${\cal V}_3$, were considered as a
simplification where the condition (\ref{E.4}) is still too restrictive
(three partial differential equations for two unknowns: $u^1(x)$ and $u^2(x)$%
). Then, the vanishing of shear: 
\begin{equation}  \label{E.5}
{\sigma_{\alpha\beta} \equiv \Sigma_{\alpha\beta} - \frac12 \Sigma^\mu_\mu
\, \hat g_{\alpha\beta} \, ; \qquad \alpha,\beta = 0,1,2 \, .}
\end{equation}
was advanced as a candidate to substitute the condition of Born-rigidity.

The condition of vanishing shear (or, equivalently, conformal rigidity)
reads: 
\begin{equation}  \label{E.5a}
\sigma_{\alpha \beta }=0\qquad \alpha ,\beta =0,1,2
\end{equation}
and yields two independent partial differential equations. Indeed, the six
equations (\ref{E.5a}) are constrained by four relations: 
\[
g^{\mu \nu }\sigma_{\mu \nu }=0\qquad {\rm and}\qquad \sigma_{\alpha \mu
}u^\mu =0,\quad \alpha =0,1,2 \,. 
\]
Since the number of unknown functions is also two, the condition (\ref{E.5a}%
) can be dealt by standard methods of solution of partial differential
systems. Namely, a non-characteristic surface ${\cal S}_2\subset {\cal V}_3$
and Cauchy data on it must be given so that a unique solution of (\ref{E.5a}%
) in a neighbourhood of ${\cal S}_2$ is determined. The surface ${\cal S}_2$
could be, for instance, a 1-parameter subcongruence.

In general, the amount of Cauchy data is much larger than our desideratum,
namely, one worldline and the vorticity of the congruence in that line. We
have however derived a way of getting a congruence out of a part of it.

In the particular case that one of the worldlines in the congruence is a
geodesic, and the (2+1)-spacetime is flat, reference \cite{LLOSA97} goes a
little further: given the congruence's vorticity on the geodesic and
assuming that strain vanishes on that worldline\footnote{%
This condition has been introduced in reference \cite{BEL94} as an
enhancement of Einstein's equivalence principle and named {\it geodesic
equivalence principle} \cite{LLOSA97}.}, the conformal rigidity condition (%
\ref{E.5a}) then determines a unique 2-parameter congruence. The latter
would be useful to model a disk whose center is at rest (or in uniform
motion), that spins at an arbitrary angular speed, and that remains {\it as
rigid as possible.}

Another remarkable result in \cite{LLOSA97} is that it exists a flat, rigid,
spatial metric, $\overline{g}_{\alpha \beta }$, which is conformal to the
Fermat tensor, $\widehat{g}_{\alpha \beta }$.

The fact that the class of {\it conformally rigid} congruences in a
(2+1)-spacetime is ``wide enough'' recalls the well know Gauss theorem \cite
{EISENHART1}:

\begin{quote}
\noindent Any Riemannian 2-dimensional space can be conformally mapped into
a flat space
\end{quote}

This suggests us a way to extend the results derived in (2+1)-spacetimes to
(3+1)-spacetimes, namely, to inspire the formulation of ``meta"-rigidity%
\footnote{%
The word ``meta"-rigidity was coined in \cite{BEL95a} to generically refer
to any relativistic extension of the notion of rigidity} conditions in some
extension of Gauss theorem to Riemannian 3-manifolds. One instance of the
latter is \cite{BEL98}, where Walberer's theorem \cite{WALB33} is taken as
the starting point. In the present paper we shall consider the following

\begin{theorem}
{\bf (Riemann)} {\rm \cite{CARTAN}} Let (M, g) be a Riemannian 3-manifold.
They exist local charts of mutually orthogonal coordinates. Moreover, this
can be done in a number of ways.
\end{theorem}

This means that six functions, namely, three coordinates $y^i$ and three
factors $f_i$, $i=1,2,3$ can be locally found such that the metric
coefficients in this local coordinates are 
\begin{equation}  \label{E.5c}
g_{ij}(y)= f^2_i(y)\delta_{ij}
\end{equation}
that is, the Riemannian metric locally admits an orthogonal basis which is
holonomic.

This result suggests us the following

\begin{definition}
A congruence is said to be Fermat-holonomic iff its Fermat tensor $\widehat{g%
}_{\alpha \beta }$ admits an orthogonal basis which is holonomic.
\end{definition}

That is, six functions exist: $y^i(x),$ $f_i(x),$ $i=1,2,3$ such that : 
\begin{equation}  \label{E.5d}
\widehat{g}_{\mu \nu }(x)dx^\mu \otimes dx^\nu =f^2_i(x) {\delta}%
_{ij}dy^i\otimes dy^j
\end{equation}
the summation convention is understood throughout the paper unless the
contrary is explicitly indicated (if one of the repeated indices is in
brackets, the convention is suspended in that formula). Greek indices run
from 1 to 4 and lattin indices from 1 to 3.

Section 2 is devoted to develop some geometrical properties of
Fermat-holonomic congruences, and in section 3 the existence of these
congruence is discussed and posed as a Cauchy problem for a partial
differential system. The method is somewhat similar to that used in proving
the existence of orthogonal triples of coordinates in a Riemannian
3-manifold (it has been specially inspiring the reading of reference \cite
{DETURK84}). In section 4 a given 2-congruence is used as the Cauchy
hypersurface for the aforementioned partial differential system, and the
problem of getting a Fermat-holonomic congruence out of one of its parts
(namely, a 2-parameter subcongruence) is studied. The kinematical meaning of
the Cauchy data is also analysed. We must finally insist in the purely local
validity of the results here derived. No global aspect of spacetimes has
been considered.

\section{Fermat-holonomic congruences}

Let ${\cal C}$ be a Fermat-holonomic 3-parameter congruence and let $u(x)$
be the unit tangent vector. According to Definition 1, the Fermat tensor, $%
\widehat g$, can be written as in equation (\ref{E.5d}). Consider the
differential 1-forms: 
\begin{equation}  \label{E6a}
\omega ^i=f_{(i)}dy^i
\end{equation}
We thus have: 
\begin{equation}  \label{E6b}
\widehat{g}={\delta }_{ij}\omega ^i{\otimes \omega }^j
\end{equation}
Moreover, since $\widehat{g}$ is orthogonal to $u$, we have that 
\begin{equation}  \label{E6}
{i(u)\omega ^l=0}
\end{equation}
As a consequence, the functions $y^i$ are constant along any worldline in
the congruence: 
\[
u(y^i)=0 
\]
Let us now introduce the differential 1-form: 
\[
\omega ^4\equiv -g(u,\_)=u_\alpha (x)dx^\alpha \,.
\]
From (\ref{E.3}) and (\ref{E6b}) it follows that 
\begin{equation}  \label{E7}
g=\widehat{g}-\omega ^4\otimes \omega ^4\equiv \eta_{\alpha \beta }\omega
^\alpha \omega ^\beta
\end{equation}

By definition [equation (\ref{E6a})] the 1-forms $\omega^i$ must be
integrable or, equivalently, they must satisfy: 
\begin{equation}  \label{E8}
d\omega^i\wedge \omega^i=0
\end{equation}

As a result we have thus proved the following

\begin{theorem}
Let (${\cal V}_4$, g) be a spacetime, ${\cal C}$ a Fermat-holonomic
3-parameter congruence and $u$ the unit velocity vector. There exist three
integrable 1-forms $\omega^i$, i = 1, 2, 3 such that completed with $%
\omega^4\equiv -g(u,\_)$, yield a $g$-orthonormal basis.
\end{theorem}

The converse theorem can be easily proved too.

\begin{theorem}
If \{$\omega^\alpha $\}$_{\alpha =1..4}$ is a $g$-orthonormal thetrad such
that $\omega^i$, $i=1,2,3$ are spacelike and integrable, then the integral
curves of $u$, the vector that results from raising the index in $\omega^4$,
form a 3-parameter Fermat-holonomic congruence.
\end{theorem}

We shall now present some geometric properties of Fermat-holonomic
congruences.

\begin{proposition}
Let $\omega^l\in \Lambda^1({\cal V}_4)$, $l=1,2,3$, be the orthonormal set
fulfilling conditions (\ref{E6}), (\ref{E7}) and (\ref{E8}) above. Then 
\begin{equation}  \label{E9}
{\cal L}(u)\omega^l\wedge \omega^l=0
\end{equation}
\end{proposition}

\smallskip\noindent {\bf Proof:} Using (\ref{E6}) and (\ref{E8}) we can
write: 
\begin{eqnarray}
{\cal L}(u)\omega^l\wedge \omega^l =
[i(u)d\omega^l+d(i(u)\omega^l)]\wedge \omega^l=i(u)d\omega^l\wedge \omega^l 
\nonumber \\
= i(u)[d\omega^l\wedge \omega^l]-d\omega^l\wedge i(u)\omega^l=0
\end{eqnarray}

\begin{proposition}
The strain rate tensor $\Sigma \equiv {\cal L}(u)\widehat{g}$ has $\omega^i, 
$ $i=1,2,3$ as principal directions. Furthermore, the same holds for any of
its Lie derivatives along the congruence: $\Sigma^{(n)}\equiv {\cal L}%
(u)^n\Sigma ={\cal L}(u)^{n+1}\widehat{g}$
\end{proposition}

\smallskip\noindent {\bf Proof:} As a consequence of proposition 1, they
exist three functions $\phi_l,$ $l=1,2,3$ such that ${\cal L}%
(u)\omega^l=\phi_{(l)}\omega^l$. Hence 
\[
\Sigma \equiv {\cal L}(u)\widehat{g}=2\phi_i\delta_{ij}\omega^i\otimes
\omega^j \,. 
\]
The second statement, concerning $\Sigma^{(n)}\,$, is easily shown by
induction. \hfill$\Box$

A sort of converse result is the following

\begin{proposition}
If ${\cal L}(u)\Sigma $ and $\Sigma $ diagonalize in the same $g$%
-orthonormal basis, then it exists an orthonormal set $\omega^l\in \Lambda^1(%
{\cal V}_4)$, such that 
\begin{equation}  \label{E9b}
{\cal L}(u)\omega^l\wedge \omega^l=0
\end{equation}
\end{proposition}

\smallskip\noindent
{\bf Proof:}
According to the hypothesis there exist three 1-forms $\rho^i,\;i=1,2,3$ such that 
\begin{eqnarray}
\label{E10a}
\widehat{g}=\delta_{ij}\rho^i\otimes \rho^j \,, \qquad &
\Sigma =2\phi_i\delta_{ij}\rho^i\otimes \rho^j  \\
\label{E10b}
  &{\cal L}(u)\Sigma =2\psi_i\delta_{ij}\rho^i\otimes \rho^j 
\end{eqnarray}
These $\rho^i$'s are orthogonal to $u$, and the same holds for ${\cal L}%
(u)\rho^l$, hence: 
\begin{equation}  \label{E10c}
{\cal L}(u)\rho^j=A_{\;k}^j\rho^k
\end{equation}

The latter can be used to calculate the Lie derivatives of (\ref{E10a}): 
\begin{eqnarray}
\Sigma = {\cal L}(u)\widehat{g}=(A_{\;i}^r\delta_{rj}+A_{\;j}^r\delta_{ri})
\rho^i\otimes\rho^j  \nonumber \\
{\cal L}(u)\Sigma =2(\dot{\phi}_i\delta_{ij}+\phi_i\delta_{ir}
A_{\;j}^r+\phi_j\delta_{jr}A_{\;i}^r)\rho^i\otimes \rho^j  \nonumber
\end{eqnarray}
which compared with (\ref{E10a}) and (\ref{E10b}) yield: 
\begin{eqnarray}
\label{E11a}
(A_{\;i}^j+A_{\;j}^i) = 2\phi_{(i)}\delta_{ij}  \\
\label{E11b}
\dot{\phi }_{(i)}\delta_{ij}+\phi_{(i)}A_{\;j}^i+
\phi_{(j)}A_{\;i}^j) = \psi_{(i)}\delta_{ij} 
\end{eqnarray}

From (\ref{E11a}) we have that: 
\begin{equation}
\label{E12}A_{\;i}^i=\phi_i\qquad ,\qquad A_{\;i}^j=-A_{\;j}^i,\quad i\neq j 
\end{equation}
which substituted in equation (\ref{E11b}) yields:
\begin{eqnarray}
\mbox{for } i=j: &\qquad \dot{\phi_i}+2\phi_i^2=\psi_i  \nonumber \\
\label{E13}
\mbox{for } i\neq j: &\qquad (\phi_{(i)}-\phi_{(j)})A_{\;i}^j=0 
\end{eqnarray}

Now three cases must be considered according to the degeneracy of the
eigenvalues of $\Sigma $.
\begin{list}
{\alph{llista})}{\usecounter{llista}}
\item In the case $\phi_1\neq \phi_2\neq \phi_3$ the set $\{\rho^i\}$ is
      unambiguously defined. Eq.(\ref{E13}) implies $A_{\;i}^j=0,i\neq j.$
      Then taking  
     (\ref{E12}) and (\ref{E10c}) into account we obtain ${\cal L}(u)\rho^i=
     \phi_{(i)}\rho^i$ and equation (\ref{E9b}) follows for $\omega^i=\rho^i$.
\item {In the case $\phi_1=\phi_2\neq \phi_3$ from (\ref{E12}) and
(\ref{E13}), we obtain:
\begin{equation}
\label{E14a}
\left.
\begin{array}{ll}
A_{\;i}^i=\phi_i \,,\qquad & i=1,2,3 \,; \\
A_{\;2}^1=-A_{\;1}^2  & 
A_{\;a}^3=-A_{\;3}^a=0 \,,\qquad a=1,2 
\end{array}
\right\}
\end{equation}
which introduced in (\ref{E10c}) yield: 
\begin{equation}
\label{E14b}
{\cal L}(u)\rho^3=\phi_{(3)}\rho^3\qquad ;\qquad {\cal L}
(u)\rho^a=A_{\;b}^a\rho^b\qquad a,b=1,2 
\end{equation}

In the present case, however, the set 
$$
\omega^3=\rho^3 \quad \mbox{and }\quad \omega^a=R_{\;b}^a\rho^b\qquad a,b=1,2 
$$
with $(R_{\;b}^a)\in O(2)$ is also a set of eigenvectors for 
$\Sigma $. Now, an orthogonal matrix $(R_{\;b}^a)$ can be found such that
${\cal L}(u)\omega^a=\phi \omega^a$, with $\phi_1=\phi_2=\phi $. 
Indeed, since 
$$
{\cal L}(u)\omega^a=[{\cal L}(u)R_{\;b}^a(R^{-1})_{\;c}^b+
(RAR^{-1})_{\;c}^a]\omega^c 
$$
it is enough to require 
$${\cal L}(u)R_{\;b}^a=R_{\;c}^a(-A_{\;b}^c+\phi \delta_{\;b}^c)$$ 
which has many solutions $(R_{\;b}^a)\in O(2)$ because, by equation (\ref{E14a}),
$-A_{\;b}^c+\phi\,\delta_b{\;}^c\,$ is skewsymmetric.}
\item The completely degenerate case $\phi_1=\phi_2=\phi_3$ can be handled
in a similar way as case (b).\hfill$\Box$
\end{list}

\begin{theorem}
${\cal L}(u)\Sigma $ and $\Sigma $ diagonalize in a common $g$-orthonormal
basis if, and only if, three functions $A$, $B$ and $C$, exist such that 
\begin{equation}
\label{E15}{\cal L}(u)\Sigma =A\widehat{g}+B\Sigma +C\Sigma^2 
\end{equation}
where $\Sigma_{\alpha \beta }^2\equiv\Sigma_{\alpha \mu
}^{}\Sigma_{\;\beta}^\mu $.  Moreover if two among the eigenvalues of
$\Sigma $ are equal, then $C=0$ can be taken, and in the completely
degenerate case, $B=C=0$ can be taken.
\end{theorem}

\smallskip\noindent
{\bf Proof:} \\
{\boldmath $(\Rightarrow)$:}\hspace*{.5em} 
Since $u$ is orthogonal to both $\Sigma $ and ${\cal L}(u)\Sigma $, 
we shall have that $\omega^4\equiv -g(u,\_)$ is in the common orthogonal 
basis $\{\omega^\alpha \}$. Thus, expressions similar
to (\ref{E10a}) and (\ref{E10b}) hold. Hence to prove (\ref{E15}) amounts to
solve the linear system: 
\begin{equation}
\label{E16}A+2\phi_iB+4\phi_i^2C=2\psi_i 
\end{equation}
for the unknowns $A$, $B$ and $C$. The determinant is:
$$
\Delta =8(\phi_2-\phi_1)(\phi_2-\phi_3)(\phi_3-\phi_1) \,.
$$
In the non-degenerate case $\Delta \neq 0$ and (\ref{E16}) has a unique
solution.

If $\phi_1=\phi_2\neq \phi_3$, only the equations for $l=2$ and $3$ in 
(\ref{E16}) are independent, there are infinitely many solutions and $C$ can
be arbitrarily chosen. In particular, $C=0$.

Finally, in the completely degenerate case, (\ref{E16}) has rank 1, hence
it admits infinitely many solutions, and $B$ and $C$ are arbitrary.

\smallskip\noindent
 
{\boldmath $(\Leftarrow)$:}\hspace*{.5em}
Assume that (\ref{E15}) holds. Since $\Sigma $ diagonalize in a
$g$-orthonormal basis, we substitute (\ref{E10a}) in (\ref{E15}) and it
follows immediately that ${\cal L}(u)\Sigma $ diagonalize in the 
same $g$-orthonormal basis.

\begin{theorem}
If $\Sigma $, ${\cal L}(u)\Sigma $ and ${\cal L}(u)^2\Sigma $ diagonalize in
a common $g$-orthonormal basis, then $\widehat{g}$, $\Sigma $, 
${\cal L}(u)\Sigma $ and ${\cal L}(u)^2\Sigma $ are linearly dependent.
(Hence,  the congruence is non-generic \cite{BEL91}.)  
\end{theorem}

\smallskip\noindent
{\bf Proof:}
By theorem 4, they exist $A$, $B$ and $C$ such that (\ref{E15})
holds. taking the Lie derivative on both sides, using that 
$\Sigma ={\cal L}(u)\widehat{g}$ and equation (\ref{E15}) itself, and taking into
account that the minimal polynomial for $\Sigma_\beta^\alpha \,$ has at
most degree 3, we arrive at: 
\begin{equation}
\label{E17}{\cal L}(u)^2\Sigma =A^{\prime }\widehat{g}+B^{\prime }\Sigma
+C^{\prime }\Sigma^2 
\end{equation}
where $A^{\prime }$, $B^{\prime }$ and $C^{\prime }$ are some suitable
functions.

If $C=0$, then (\ref{E15}) already proves the theorem.

If, on the contrary $C\neq 0$, we can derive $\Sigma^2$ from (\ref{E15})
and substitute it into (\ref{E17}), so arriving at:
$$
{\cal L}(u)^2\Sigma =\left(A^{\prime }-\frac AC\right) \,\widehat{g}
+\left(B^{\prime }-\frac BC\right)\,\Sigma +
\frac{C^{\prime }}C \,{\cal L}(u)\Sigma 
$$
which ends the proof. \hfill$\Box$

\section{Existence of Fermat-holonomic congruences}
According to the theorems 2 and 3 in section 2, proving the existence of
Fermat-holonomic congruences is equivalent to prove the existence of a
$g$-orthonormal basis $\{\omega^\alpha \}$ such that:

\begin{list}
{(\roman{llista})}{\usecounter{llista}}
\item \quad $\omega^i$ is spacelike, and
\item \quad $d\omega^i\wedge \omega^i=0$ \hfill (\ref{E8})
\end{list}
The dual thetrad will be denoted $\{e_\alpha \}$ and the commutation
relations: 
\begin{equation}
\label{e18}
[e_\alpha ,e_\beta ]=\,C_{\alpha \beta }^\mu e_\mu \qquad 
d\omega^\alpha =-\frac 12\,C_{\mu \nu }^\alpha \,\omega^\mu \wedge \omega^\nu 
\end{equation}
using the latter, (\ref{E8}) can be written as: 
\begin{equation}
\label{e19}-\frac 12\,C_{\mu \nu }^i\omega^\mu \wedge \omega^\nu \wedge
\omega^i=0 \qquad i=1,2,3
\end{equation}
which in turn is equivalent to: 
\begin{equation}
\label{e19b}
\left.
\begin{array}{ll}
C_{jk}^i=0\qquad & i\neq j\neq k \\ 
C_{4j}^i=0\qquad & i\neq j 
\end{array}
\right\}
\end{equation}

Taking into account the relationship between $C_{\alpha \beta }^\gamma $ and
the Riemannian connexion coefficients in an orthonormal frame
\cite{CHOQUET}, equations (\ref{e19b}) are equivalent to:  
\begin{eqnarray}
\label{e20a}
 \gamma_{jk}^i  =0  & \qquad  i\neq j\neq k \\
\label{e20b}
C_{4j}^i  =\gamma_{4j}^i-\gamma_{j4}^i=0 & \qquad   i\neq j 
\end{eqnarray}

For a Riemannian connexion in an orthonormal frame, it holds:
$$
\gamma_{\alpha k}^i=-\gamma_{\alpha i}^k\qquad ,\qquad \gamma_{\alpha
k}^4=\gamma_{\alpha 4}^k\qquad \alpha =1...4,\quad i,k=1,2,3 
$$
where the signature $(+++-)$ has been used.
Hence, at most three among the equations (\ref{e20a}) are independent.

Since 
\begin{equation}
\label{e22}\gamma_{\alpha \beta }^\gamma =(\omega^\gamma ,e_\alpha^{\
\nabla }e_\beta )=\eta^{\gamma \nu }g(e_\nu ,e_\alpha^{\ \nabla }e_\beta ) 
\end{equation}
equations (\ref{e20a}) and (\ref{e20b}) yield a first-order partial
differential system of nine equations where the unknown is the orthonormal
frame $\{e_\alpha \}$.

\subsection{The Cauchy problem}
Given a hypersurface ${\cal S}_0$, consider a 1-parameter family of
hypersurfaces, ${\cal S}_\lambda $, containing it. Let $n$ be the unit
orthogonal vector field which, by construction, is hypersurface
orthogonal, i. e., $\nu =g(n,\_)$ is an integrable 1-form.

Relatively to $n$, each vector $e_\alpha $ of the sought frame can be
decomposed in an orthogonal part $e_\alpha^\top$ (which is tangent to the
hypersurfaces ${\cal S}_\lambda$) and a parallel part, namely: 
\begin{equation}
\label{e23}
e_\alpha =e_\alpha^\top+n_\alpha n 
\end{equation}
where $n_\alpha \equiv g(e_\alpha ,n)$.

Substituting this decomposition in (\ref{e20a}) and (\ref{e20b}), taking 
into account equation (\ref{e22}) and the fact that the connexion is
Riemannian, we respectively arrive at: 
\begin{eqnarray}
\label{e24a}
\gamma_{jk}^i = n_j W_{ik}+
 g(e_i,e_j^{\top\nabla }e_k)=0 &\qquad i\neq j\neq k  \\
\label{e24b}
C_{4k}^i = n_4W_{ik}-n_kW_{i4}+H_{ik}=0 &\qquad
i\neq k 
\end{eqnarray}
where 
\begin{equation}
\label{e25a}
W_{\alpha \beta }\equiv g(e_\alpha ,n^{\nabla }e_\beta ) 
\end{equation}
is skewsymmetric, and 
\begin{equation}
\label{e25}
H_{ik}\equiv 
g(e_i ,[e_4^{\top},e_k^\top])+
n_i \left(e_4^{\top}(n_k) - e_k^{\top}(n_4)\right)
+ n_k g(e_i ,e_4^{\top\nabla }n)-n_4 g(e_i ,e_k^{\top\nabla }n) 
\end{equation}
only depends on the unknowns and their derivatives along directions that are
tangential to ${\cal S}_\lambda $.

If, and only if, $n_j\neq 0$, the nine equations (\ref{e24a}) and (\ref{e24b}) 
can be solved for the six independent components of $W_{\alpha
\beta }$. A straight manipulation yields: 
\begin{eqnarray}
\label{e26a}
W_{ik} = -\frac 1{n_j}g(e_i,e_j^{\top\nabla }e_k^{})\qquad
 \qquad\qquad  &i\neq j\neq k  \\
\label{e26b}
W_{i4} = \frac 1{n_k}H_{ik}-\frac{n_4}{n_k n_j}
g(e_i,e_j^{\top\nabla }e_k^{}) &\qquad i\neq j\neq k 
\end{eqnarray}

Notice that for each value of $i=1,2,3$ there are two ways of choosing 
$j\neq k$ in equation (\ref{e26b}). This comes from the fact that the system 
(\ref{e24a})--(\ref{e24b}) is overdetermined. Hence, three subsidiary
conditions follow: 
\begin{equation}
\label{e27}
S_i\equiv \frac 1{n_k}H_{ik}-\frac 1{n_j}H_{ij}-
\frac{n_4}{n_kn_j}g(e_i,e_j^{\top\nabla }e_k^{}-e_k^{\top\nabla
}e_j^{})=0\qquad i\neq j\neq k 
\end{equation}
these conditions only depend on ``tangential derivatives'' of the unknowns.

\begin{lemma} \label{L1}
Given an orthonormal thetrad $\{\overline{e}_\alpha \}$ on ${\cal S}_0$,
such that $g(n,\overline{e}_i)\neq 0$, it exists a neighbourhood ${\cal U}$ of 
${\cal S}_0$ and an orthonormal thetrad $\{e_\alpha \}$ which is a solution
of 
\begin{equation}
\label{e28a}
\left.
\begin{array}{l}
\displaystyle{W_{ik}= -\frac 1{n_j}g(e_i,e_j^{\top\nabla }e_k)} \\ 
\displaystyle{W_{\ i4}=\frac 1{n_k}H_{ik}-\frac{n_4}{n_kn_j}}
  g(e_i,e_j^{\top\nabla}e_k^{}) 
\end{array} \right\}
\end{equation}
[where $(ikj)$ is a cyclic permutation of $(123)$] and such that $e_\alpha =
\overline{e}_\alpha $ on ${\cal S}_0$.
\end{lemma}

\smallskip\noindent
{\bf Proof:}
Let $\{\widetilde{e}_\alpha \}$ be a given orthonormal frame. In
terms of it the unknown frame $\{e_\alpha \}$ can be written as:
$$
e_\alpha =L_{\ \alpha }^\mu \widetilde{e}_\mu 
$$
where $(L_{\ \alpha }^\beta)$ is a Lorentz matrix valued function.

Substituting the latter into equation (\ref{e25a}) we have:
$$
W_{\alpha \beta }=L_{\ \alpha }^\mu g(\widetilde{e}_\mu ,n^{\nabla
}(L_{\ \beta }^\nu \widetilde{e}_\nu )) \,.
$$
Then, after a straightforward calculation, we arrive at: 
\begin{equation}
\label{e28}
n L_{\ \alpha }^\beta =L_{\ }^{\beta \mu }W_{\mu
\alpha }-\eta^{\beta \mu }L_{\ \alpha }^\rho 
g(\widetilde{e}_\mu ,n^{\nabla }\widetilde{e}_\rho ) 
\end{equation}
where indices are raised by contraction with $\eta^{\alpha \beta }$ and
$nL_{\ \alpha}^beta$ is the directional derivative along $n$.

The second term in the right hand side is known and, since 
$g(n,\widetilde{e}_i)\neq 0$ on 
${\cal S}_0$, equation (\ref{e28a}) yield $W_{\alpha \beta }$
on ${\cal S}_0$ as a function of $e_\alpha $ and their ``tangential''
derivatives (i. e. L$_{\ \alpha }^\beta $ and their ``tangential''
derivatives). Hence, the Cauchy-Kowalevski theorem \cite{JOHN71} can be
invoked to end the proof. \hfill $\Box$

\begin{lemma} \label{L2}
Let $\{e_\alpha \}$ be an orthonormal thetrad, which is a solution 
of (\ref{e28a}) and fulfills the subsidiary conditions (\ref{e27}) on
${\cal S}_0$. Then (\ref{e27}) holds in the neighbourhood ${\cal U}$ of
${\cal S}_0$ where $\{e_\alpha \}$ is defined.
\end{lemma}

\smallskip\noindent
{\bf Proof:}
Since $\{e_\alpha \}$ is a solution of (\ref{e28a}), 
\begin{equation}
\label{e29}C_{jk}^i=0\qquad ,\qquad C_{4k}^i=0\qquad C_{4j}^i=- n_j S_i 
\end{equation}
where $(ikj)$ is a cyclic permutation of $(123)$ and $S_i$ is defined in
(\ref{e27}).

The commutation coefficients $C_{\alpha \beta }^\mu $ satisfy the Jacobi like
identity: 
\begin{equation}
\label{e30}
\eta^{\alpha \rho \mu \nu }(e_\rho C_{\mu \nu }^\beta -C_{\sigma
\rho }^\beta C_{\mu \nu }^\sigma )=0 
\end{equation}

Taking $\alpha =\beta =i$ in (\ref{e30}) and using (\ref{e29}) we arrive at:
$$
e_kC_{4j}^i+C_{4j}^i(C_{k(j)}^{(j)}+C_{k4}^4-C_{k(i)}^{(i)})=0,\qquad \varepsilon
_{ikj}=1 
$$
and taking the decomposition (\ref{e23}) into account, and the fact that 
$n_k\neq 0$ we have 
\begin{equation}
\label{e31}
n\left(C_{4j}^i\right)+\frac 1{n_k}[e_k^\top C_{4j}^i+A_kC_{4j}^i]=0, 
\end{equation}
with $A_k=C_{k(j)}^{(j)}+C_{k4}^4-C_{k(i)}^{(i)}$.

The latter can be considered as a partial differential system on $C_{4j}^i$.
In case that $n_k\neq 0$, the hypersurface ${\cal S}_0$ is
non-characteristic, and the Cauchy-Kowalevski theorem states that (\ref{e31}) 
has a unique solution, which for $S_i=0$ on ${\cal S}_0$, ensures that
$C_{4j}^i=0 $ in a neighbourhood of ${\cal S}_0\,$ and, equivalently,
$S_i=0$ in a neighbourhood of ${\cal S}_0$.

Summarizing, we have shown the following 

\begin{theorem} \label{T6}
Given an orthonormal thetrad $\{\overline{e}_\alpha \}$ on ${\cal S}_0$,
such that:
\begin{list}
{(\roman{llista})}{\usecounter{llista}}
\item $\qquad g(n,\overline{e}_i)\neq 0\quad $ and
\item $\qquad S_i=0\quad $ on ${\cal S}_0$
\end{list}
it exists an orthonormal frame $\{e_\alpha \}$ in a neighbourhood ${\cal U}$
of ${\cal S}_0$, such that:
\begin{list}
{(\roman{llista})}{\usecounter{llista}}
\item it is a solution of (\ref{e28a}) and (\ref{e27}), and
\item $\quad e_\alpha =\overline{e}_\alpha \quad $ on $\,{\cal S}_0$
\end{list}
Thus, the congruence generated by $u=e_4$ is Fermat-holonomic.
\end{theorem}

The proof is straightforward from lemmas \ref{L1} and \ref{L2}. \hfill $\Box$

\section{Getting a Fermat-holonomic 3-congruence out of a given 2-congruence}
Let ${\cal S}_0\,$be the hypersurface spanned by a given 2-congruence of
worldlines and let $\overline{u}$ be the unit velocity vector.
(Hereafter, a bar over a symbol indicates that we are only considering the
values of that object on ${\cal S}_0$.)

\subsection{Are the subsidiary conditions consistent?}
We are interested in completing an orthonormal thetrad
$\{\overline{e}_\alpha \}$ on ${\cal S}_0$, such that: 
$\overline{e}_4=\overline{u}$, \,$\overline{n}_i\equiv
g(\overline{e}_i,\overline{n})\neq 0\,$, and that $S_i=0$ on ${\cal
S}_0$, where $\overline{n}$ is the unit vector orthogonal 
to ${\cal S}_0.$

For each $\overline{e}_i$, we consider the decomposition (\ref{e23}):
$$
\overline{e}_i=\overline{e}_i^\top+\overline{n}_i\overline{n} 
$$
and take the unit vector 
\begin{equation}
\label{e34}\widehat{a}_i\equiv \frac{\overline{e}_i^\top}{\left| 
\overline{e}_i^\top\right| } 
\end{equation}

Furthermore, we can consider the combinations: 
\begin{equation}
\label{e35}
\overline{b}_i=\epsilon_i^{\ jk}\overline{n}_j\,\overline{e}_k \qquad
\qquad i=1,2,3 
\end{equation}
and define the unit vector $\widehat{b}_i\equiv 
\overline{b}_i/|\overline{b_i}|$.
It is straightforward to see that $\{\overline{u},\overline{n},\widehat{b}_i,
\widehat{a}_i\}$ is an orthonormal thetrad at any point in ${\cal S}_0$.

We also have that 
$$g(\widehat{a}_i,\widehat{a}_j)=\frac 1{\left| 
\overline{e}_i^\top\right| \left| \overline{e}_j^\top\right|}
g(\overline{e}_i^\top, \overline{e}_j^\top)=\frac 1{\left|
\overline{e}_i^\top\right| \left| \overline{e}_j^\top\right|
}(\delta_{ij}-\overline{n}_i\overline{n}_j)$$ 
Now, since $\overline{n}_i\neq 0$, then $\left| \overline{n}_j\right| <1$ and
it follows that:
\begin{equation}
\label{e36}
\forall i\neq j\left| g(\widehat{a}_i,\widehat{a}_j)\right| <1\qquad 
{\rm and} \qquad 
g(\widehat{a}_i,\widehat{a}_j)\neq 0 
\end{equation}

Thus, for any triad $\{\overline{e}_1,\overline{e}_2,\overline{e}_3\}$ in 
$T_x{\cal V}_4$, $x \in{\cal S}_0$, such that: 
$g(\overline{e}_i,\overline{u})=0$, and 
$g(\overline{e}_i,\overline{n})=\overline{n}_i\neq 0$, we can obtain a triad 
$\widehat{a}_i\in $ $T_x{\cal S}_0$, $i=1,2,3$, such that
$g(\widehat{a}_i,\overline{u}) =0$ and that (\ref{e36}) holds.

The converse result is the following

\begin{theorem}
Given a triad $\widehat{a}_i\in $ $T_x{\cal S}_0$, $i=1,2,3$, 
such that $g(\widehat{a}_i,\overline{u})=0$ and fulfills (\ref{e36}), 
a triad $\{\overline{e}_i\}$ can be obtained such that
$g(\overline{e}_i,\overline{n})\neq 0\,$ and that
$\{\overline{e}_1,\overline{e}_2,\overline{e}_3,\overline{e}_4=
\overline{u}\}$ is an orthonormal frame.
\end{theorem}

\smallskip\noindent
{\bf Proof:}
We must find $A_i$ and $B_i$ such that 
$$
\overline{e}_i=A_i\widehat{a}_i+B_i\overline{n} 
$$
since $\widehat{a}_i$, and $\overline{n}$ are $u$-orthogonal, so will be 
$\overline{e}_i$.

Now, the condition $g(\overline{e}_i,\overline{e}_j)=\delta_{ij}$ implies: 
\begin{equation}
\label{e36a}A_i^2+B_i^2=1 
\end{equation}
\begin{equation}
\label{e36b}A_iA_jg(\widehat{a}_i,\widehat{a}_j)+B_iB_j=0\qquad i\neq j 
\end{equation}
From which we easily obtain: 
\begin{equation}
\label{e37}B_i=A_i\sqrt{-\frac{g(\widehat{a}_i,\widehat{a}_j)\dot 
g(\widehat{a}_i,\widehat{a}_k)}{g(\widehat{a}_j,\widehat{a}_k)}}\qquad i\neq
j\neq k 
\end{equation}
and 
\begin{equation}
\label{e38}A_i=\sqrt{\frac{g(\widehat{a}_j,\widehat{a}_k)}{g(\widehat{a}_j,
\widehat{a}_k)-g(\widehat{a}_i,\widehat{a}_j)\cdot g(\widehat{a}_i,
\widehat{a}_k)}}\qquad i\neq j\neq k 
\end{equation}
The right hand side of (\ref{e37}) is well defined because, by the
hypothesis, the inequalities (\ref{e36}) hold.

The denominator in (\ref{e38}) neither vanishes, as a consequence of 
(\ref{e36}) too. Indeed, denoting by 
$\varphi_{ij}\,$ the angle between $\widehat{a}_i$ and $\widehat{a}_j$, and
taking into account that $\left| \varphi_{ij}\right| +\left| \varphi
_{jk}\right| +\left| \varphi_{ki}\right| =2\pi$, $\, i\neq j\neq k\,$
this denominator reads:
$$
\cos (\varphi_{jk})-\cos (\varphi_{ij})\cdot \cos (\varphi_{ik}) =
\sin (\varphi_{ij})\cdot \sin (\varphi_{ik}) \qquad i\neq j\neq k 
$$
which does not vanish because $\left| g(\widehat{a}_i,\widehat{a}_j)\right|
=\left| \cos (\varphi_{ij})\right| <1\,,\quad \forall i\neq j$. 
\hfill$\Box$

We shall attempt to find $\widehat{a}_1,\widehat{a}_2,\widehat{a}_3\,$ 
such that the triad $\{\overline{e}_i\}$\thinspace so reconstructed fulfill
the subsidiary condition (\ref{e27}). 
It is easily seen that $S_i$ can be written as:
$$
S_i\equiv\frac 1{n_j}g({e}_i,[{e}_4,{e}_j])- \frac 1{n_k}g({e}_i,[{e}_4,{e}_k])
=- \frac 1{{n}_j\,{n_k}}\,g({e}_i,[{e}_4,b_i]) 
$$
(wher $(ikj)$ is a cyclic permutation of $(123)$) 
on the hypersurface ${\cal S}_0$, that is:
$$
\overline{S}_i=-\frac 1{\overline{n}_j\overline{n}_k} \,
 g(\overline{e}_i,[\overline{u},\overline{b}_i]) 
$$

Taking now into account that $\overline{u}\,$and $\overline{b}_i\,$ are
tangent to ${\cal S}_0$, (hence, $[\overline{u},\overline{b}_i]$ is 
tangent too), we can write: 
$$
-\overline{n}_j\overline{n}_k\,\overline{S}_i=
g(\overline{e}_i^\top,[\overline{u},\overline{b}_i])=\left| 
\overline{e}_i^\top\right| \cdot \left| \overline{b}_i\right| g(\widehat{a}_i,
[\overline{u},\widehat{b}_i]) 
$$
so that the subsidiary conditions $\overline{S}_i=0$ are equivalent to 
\begin{equation}
\label{e39}g(\widehat{a}_i,[\overline{u},\widehat{b}_i]) =0
\end{equation}

Let $\overline{v}$ and $\overline{m}$ be two unit vector fields such that
$\{\overline{u}, \overline{m}, \overline{v}\}$ is an orthonormal basis on
the tangent space of ${\cal S}_0$. In terms of this basis, we can write:
\begin{equation}
\label{e39b}
\left.
\begin{array}{l}
\widehat{b}_i=\cos \theta_i\,\overline{m}+\sin \theta_i\,\overline{v} \\ 
\widehat{a}_i=-\sin \theta_i\,\overline{m}+\cos \theta_i\,\overline{v} 
\end{array}
\right\}
\end{equation}
and the conditions (\ref{e39}) lead to: 
\begin{equation}
\label{e40}
\dot\theta + \frac14 \sin 2\theta_i\, (\overline\Sigma_{vv} - 
    \overline\Sigma_{mm}) + \frac12 \cos 2\theta_i\,\overline\Sigma_{mv}
    + \frac12 \left(g(\overline{v},[\overline{u},\overline{m}]) -
               g(\overline{m},[\overline{u},\overline{v}])\right) = 0
\end{equation}
which is an ordinary differential equation for each $\theta_i$, $i=1,2,3$,
and has a solution for every initial data $\theta_i^0$ given in a
submanifold ${\cal M}\subset{\cal S}_0$, such that it is nowhere tangent to
$\overline{u}$. 

Summarizing, we have thus proved that given a 2-congruence, it spans a
hypersurface ${\cal S}_0$ on which Cauchy data $\{\overline{e}_1,
\overline{e}_2,\overline{e}_3,\overline{e}_4=\overline{u}\}$ can be found
fulfilling the subsidiary conditions\thinspace (\ref{e27}). Moreover, this
can be done in an infinite number of ways.

\subsection{The kinematical meaning of the Cauchy data}
In the particular case considered in this section, where the Cauchy
hypersurface ${\cal S}_0$ is spanned by a 2-congruence, we shall analyse
the kinematical significance of the Cauchy data
$\{\overline{e}_i\}_{i=1,2,3}$.

According to Proposition 2, the latter is a principal basis for 
$\overline\Sigma$ (i. e., the values of the strain rate of the 
Fermat-holonomic 3-congruence. We shall
see how  $\{\overline{e}_i\}_{i=1,2,3}$ determine $\overline\Sigma$.

Indeed, let $\overline{n}$ be  the unit vector normal to ${\cal S}_0$ 
and $\overline{\nu}\equiv g(\overline{n},\_)$. $\overline\Sigma$
can be written as:
\begin{equation}
\label{E59}
\overline\Sigma=\overline\Sigma^0 + \overline{\pi} \otimes\overline{\nu}
+\overline{\nu}\otimes\overline{\pi} +
\overline{s}\,\overline{\nu}\otimes\overline{\nu}
\end{equation}
where:
$$\overline\Sigma^0(\overline{n},\_) =\overline\Sigma^0(\_,\overline{n}) =0 
\qquad {\rm and} \qquad \langle\overline{\pi},\overline{n}\rangle =0 \,.$$

The condition that $\overline{e}_i$ is a principal vector then reads:
$$ \exists \lambda_i \quad\mbox{such that} \quad 
\overline\Sigma(\overline{e}_i,\_) =
\lambda_i\overline{\omega}^i $$
which, using (\ref{E59}) and considering the parallel and orthogonal
parts, respectively yields:
\begin{eqnarray}
\label{E60a}
\langle\overline{\pi},\overline{e}_i\rangle + 
\overline{s} \,\overline{n}_i =\lambda_i\overline{n}_i
\\
\label{E60b}
\overline\Sigma^0(\overline{e}_i^\top,\_) + \overline{n}_i \overline{\pi}
= \lambda_i\,\widehat{g}(\overline{e}_i^\top,\_)
\end{eqnarray}
Now, taking into account (\ref{e23})  and expressions like
$$\overline{n}^l\,\overline{e}_l = \overline{n} \qquad  \mbox{and} \qquad
\sum_{l=1}^3 (\overline{n}^l)^2 = 1 \,,$$ after a short calculation, we
arrive at:
\begin{eqnarray}
\label{E61}
\overline{p} = -\sum_{j\neq l} \frac 1{\overline{n}_j}\,
\overline\Sigma^0(\overline{e}_j^\top,\overline{e}_l^\top)
\overline{e}_l^\top
\, ,\qquad & \mbox{where}\quad \overline{\pi} = \widehat{g}(\overline{p},\_)\\
\label{E62}
\overline{s} = 2
\overline\Sigma^0(\overline{e}_j^\top,\overline{e}_l^\top)\,
\delta^{lj}+\sum_{j<l}
\frac 1{\overline{n}_j\overline{n}_l}\overline\Sigma^0(\overline{e}_j^\top,
\overline{e}_l^\top) & {}
\\
\label{E63}
\lambda_i = \overline\Sigma^0(\overline{e}_h^\top,\overline{e}_l^\top) 
\delta^{lh} + \frac 1{\overline{n}_j
\overline{n}_k}\overline\Sigma^0(\overline{e}_j^\top,\overline{e}_k^\top) 
  & {}
\end{eqnarray}

Let us now see what does $\overline\Sigma^0$ mean. 
Given any couple of vector fields, $\overline{v}$ and
$\overline{w}$, that are tangent to ${\cal S}_0$, we have that:
\begin{eqnarray}
\overline\Sigma^0(\overline{v},\overline{w})  =&
    \overline\Sigma(\overline{v},\overline{w}) = {\cal L}(u)
    \widehat{g}(\overline{v},\overline{w}) &\nonumber \\ 
= &\overline{u}\left(\widehat{g}(\overline{v},\overline{w})\right) -
     (\widehat{g}([\overline{u},\overline{v}],\overline{w}) -
     (\widehat{g}(\overline{v},[\overline{u},\overline{w}])& \nonumber
\end{eqnarray}
Now, since ${\cal S}_0$ is a submanifold,
$[\overline{u},\overline{v}]$ and $[\overline{u},\overline{w}]$ 
are also tangent to ${\cal S}_0$. Hence, the Fermat tensor $\widehat{g}$ 
on the right hand side can be replaced by its restriction to ${\cal S}_0$,
namely, $\widehat{g}^0$. So that,
$$ \overline\Sigma^0 = {\cal L}(\overline{u}) \widehat{g}^0 $$
Thus, $\overline\Sigma^0$ is the strain rate of the given 2-congruence, 
${\cal C}_2$ in the Riemannian submanifold $({\cal S}_0,\widehat{g}^0)$. 

We have so far shown the following

\begin{proposition}\label{P6}
The Cauchy data $({\cal S}_0, \overline{e}_1, \overline{e}_2,
\overline{e}_3, \overline{e}_4=\overline{u})$ determine $\overline\Sigma$,
i. e., the values on ${\cal S}_0$ of the strain rate of the Fermat-holonomic
congruence obtained as a solution of the partial differential system
(\ref{e24a})--(\ref{e24b}).  
\end{proposition}

The converse is true only if all the eigenvalues of $\overline\Sigma$
restricted to spacelike vectors have multiplicity 1. Indeed, if some
eigenvalue of $\overline\Sigma$ has multiplicity greater than 1, then the
eigenvectors are determined up to a rotation.

\subsection{Uniqueness}
From Proposition \ref{P6} it seems to follow that, given the 2-congruence 
${\cal C}_2$ and $\overline\Sigma$, there is a unique
Fermat-holonomic 3-congruence that includes ${\cal C}_2$ and
its strain rate takes the values $\overline\Sigma$ on the submanifold
${\cal S}_0$. We shall now see that, although this is the generic case, it
is not allways true. 

Once ${\cal C}_2$ is given, $\overline\Sigma$ is no more arbitrary.
Indeed, the first block $\overline\Sigma^0$ in the decomposition (\ref{E59}) 
is already determined by ${\cal C}_2$. Hence, it is enough to give the
1-form $\overline{\pi}$ and the scalar $\overline{s}$ on ${\cal S}_0$, 
and use (\ref{E59}) to determine $\overline\Sigma$.
Then, a basis $\{\overline{e}_1,\overline{e}_2,\overline{e}_3\}$ of
spacelike eigenvectors of $\overline\Sigma$ can be chosen.

In order that $({\cal S}_0,\overline{e}_1, \overline{e}_2, \overline{e}_3,
\overline{e}_4=\overline{u})$ is a suitable set of Cauchy data for the
partial differential system (\ref{e24a})--(\ref{e24b}) it is 
necessary that $\overline{n}_i\neq 0$, $i=1,2,3$. 
This is equivalent to require that no $\overline{e}_l$ is tangent to
${\cal S}_0$. 

Now, for $\overline{e}_l$ to be tangent to ${\cal S}_0$ we should have that:
$$ \langle \overline{\pi},\overline{e}_l\rangle =0 \qquad \mbox{and}  \qquad 
\overline\Sigma^0(\overline{e}_l,\_) =
\lambda_l\,\widehat{g}^0(\overline{e}_l,\_) \,,$$ 
where (\ref{E60a}) and (\ref{E60b}) have been used. That is,
$\overline{e}_l$ is a principal vector of $\overline\Sigma^0$ and
$\overline{\pi}$ is orthogonal to $\overline{e}_l$. Hence, $\overline{p}$
itself should be an eigenvector of $\overline\Sigma^0$.

Thus, avoiding to choose $\overline{p}$ among the eigenvectors of
$\overline\Sigma^0$ guarantees that $\overline{n}_i\neq 0$, $i=1,2,3$. Of
course, this is not possible when $\overline\Sigma^0$ is a multiple of
$\widehat{g}^0$. 

So that, several possibilities occur:
\begin{list}
{(\Alph{llista})}{\usecounter{llista}}
\item {Either $\overline\Sigma$ has three simple eigenvalues and therefore
     the eigenvectors $\{\overline{e}_i\}_{i=1,2,3}$ are uniquely
     determined, in which case:
     \begin{list}
     {(\Alph{llista}.\arabic{llista1})}{\usecounter{llista1}}
     \item either $\overline{n}_i\neq 0$, $i=1,2,3$, and $({\cal
            S}_0,\overline{e}_1, \overline{e}_2, \overline{e}_3,
       \overline{e}_4=\overline{u})$ is a suitable set of Cauchy data, 
     \item or some $\overline{n}_l=0$, and ${\cal S}_0$ is characteristic.
     \end{list}
     }
\item Or some eigenvalue of $\overline\Sigma$ is not simple.
     Then, the basis of eigenvectors is determined up to a rotation. 
\end{list}

Therefore, only in the case (A.1) ${\cal C}_2$ and $\overline\Sigma$
determine a unique Fermat-holonomic 3-congruence.

\section{Conclusion and outlook}
Looking for a less restrictive substitute for Born's relativistic
definition of rigid motion, we have suggested the definition of {\it
Fermat-holonomic} motion, namely, a 3-parameter congruence of timelike
worldlines which admits an adapted system of coordinates
$(t,y^1,y^2,y^3)$, such that the hypersurfaces $y^i=$constant, $i=1,2,3$,
are mutually orthogonal (i.e., the Fermat tensor is diagonal:
$\widehat{g}_{ij}(t,y) = 0$ whenever $i\neq j$).

We have expressed this condition as a partial differential system and
analysed the Cauchy problem. We have proved ---theorem \ref{T6}--- that,
given a 2-parameter congruence ${\cal C}_2$ and a triad of spatial vectors
$\{\overline{e}_i\}_{i=1,2,3}$ on ${\cal S}_0$ (the track of ${\cal
C}_2$), there is a unique Fermat-holonomic 3-congruence, 
${\cal C}_3$, containing ${\cal C}_2$ and admitting an adapted coordinate
system such that the spatial coordinate curves are tangent to
$\overline{e}_i$ on ${\cal S}_0$.

Moreover, the given directions $\{\overline{e}_i\}$ are related to the
eigenvectors of the strain rate tensor $\overline\Sigma$ and, except in
the case that the shear of ${\cal C}_3$ vanishes on ${\cal S}_0$, we have
shown that ${\cal C}_2$ together with $\overline\Sigma$ determine ${\cal
C}_3$ in a neigbourhood of ${\cal S}_0$. Hence, a Fermat-holonomic
congruence is determined by ``a part of it".

The definition of Fermat-holonomic congruences has been devised as an
extension to a (3+1)-spacetime of the shear-free congruences studied in
reference  \cite{LLOSA97} in a similar context, for the simplified problem of a
(2+1)-spacetime. There, we went further and, adding symmetry arguments and
the so called {\it geodesic equivalence principle} \cite{LLOSA97},
\cite{BEL94}, a unique
shear-free congruence was obtained out of a worldline and the angular
velocity on it, just like in the case of Newtonian rigid motions. In a
forthcoming paper we shall try to supplement in a similar way (symmetries
plus geodesic equivalence principle) the general results that have been
derived here, to model an arbitrary rotational motion with a fixed point.

\section{Acknowledgments}

One of us is indebted to M. A. Garcia Bonilla for helpful suggestions and
comments. This work is partly supported by DIGICyT, contract no.PPB96-0384
and by Institut d'Estudis Catalans (S.C.F.).




\begin{thebibliography}{9}
\bibitem{BEL90}  BEL, Ll., {\it Proc. Encuentros Relativistas Espa
oles-1989:  Recent Developments in Gravitation},  World Scientific
(Singapore,1990)

\bibitem{BORN09}  BORN, M., {\it Phys. Z.}, {\bf 10} (1909) 814

\bibitem{HER-NOE10}  HERGLOTZ, G., \AP, {\bf 31}, 393 (1910);  NOETHER, F., %
\AP, {\bf 31}, 919 (1910)

\bibitem{LLOSA97}  LLOSA, J., \CQG, {\bf 14}, 165 (1997)

\bibitem{BEL94}  BEL, Ll. MARTIN, J. and MOLINA, A., \JPSJ, {\bf 63} (1994)
4350.

\bibitem{BEL95a}  BEL, Ll. and LLOSA, J., {\it Gen. Rel. Grav.}, {\bf 27}
(1995) 1089.

\bibitem{EISENHART1}  EISENHART, L. P., {\it Riemannian Geometry},  p. 92,
Princeton University Press (Princeton, 1960).

\bibitem{BEL98}  BEL, Ll., {\it Proc. Encuentros Relativistas  Espa
oles-1998: Relativity and Gravitation in General},  World Scientific
(Singapore,1999)

\bibitem{WALB33}  WALBERER, P. {\it Hamburg. Abhandlungen}, {\bf 10} (1933)
147.

\bibitem{CARTAN}  CARTAN, E., {\it Les syst\`{e}mes diff\'{e}rentiels
ext\'{e}rieurs et leurs applications g\'{e}om\'{e}triques}, p. 187, Hermann
(Paris, 1971).

\bibitem{DETURK84}  DETURK, D. M. and YANG, D., {\it Duke Math. Jour.}, {\bf %
51} (1984) 243.

\bibitem{BEL91}  BEL, Ll. and COLL, B., {\it Gen. Rel. Grav.}, {\bf 25}
(1993) 613.

\bibitem{CHOQUET}  CHOQUET-BRUHAT, Y., DEWITT-MORETTE, C. and
DILLARD-BLEICK, M., {\it Analysis, Manifolds and Physics}, p. 308, revised
edition, North-Holland (Amsterdam, 1987).

\bibitem{JOHN71}  JOHN, F., {\it Partial Differential Equations},  p.56,
Springer (New York, 1971)
\end{thebibliography}
\end{document}